\documentclass[twocolumn,aps,superscriptaddress]{revtex4}

\usepackage{amssymb}
\usepackage{amsmath}
\usepackage{graphicx}
\usepackage[normalem]{ulem}
\usepackage[dvips]{color}
\usepackage{appendix}

\usepackage{color}
\usepackage[normalem]{ulem}

\begin{document}

\title{Elliptic flow splittings in the Polyakov-looped Nambu-Jona-Lasinio transport model}
\author{Wen-Hao Zhou}
\affiliation{Shanghai Institute of Applied Physics, Chinese Academy
of Sciences, Shanghai 201800, China}
\affiliation{University of Chinese Academy of Sciences, Beijing 100049, China}
\author{He Liu}
\affiliation{Science School, Qingdao University of Technology, Qingdao 266000, China}
\author{Feng Li}
\affiliation{School of Physical Science and Technology, Lanzhou University, Lanzhou, Gansu, 073000, China}
\author{Yi-Feng Sun}
\affiliation{Laboratori Nazionali del Sud, INFN-LNS, Via S. Sofia 62, I-95123 Catania, Italy}
\author{Jun Xu\footnote{Corresponding author: xujun@zjlab.org.cn}}
\affiliation{Shanghai Advanced Research Institute, Chinese Academy of Sciences, Shanghai 201210, China}
\affiliation{Shanghai Institute of Applied Physics, Chinese Academy of Sciences, Shanghai 201800, China}
\author{Che Ming Ko}
\affiliation{Cyclotron Institute and Department of Physics and Astronomy,
Texas A$\&$M University, College Station, Texas 77843, USA}
\date{\today}

\begin{abstract}
To incorporate the effect of gluons on the evolution dynamics of the quark matter produced in relativistic heavy-ion collisions, we extend the 3-flavor Nambu-Jona-Lasinio (NJL) transport model to include the contribution from the Polyakov loops. Imbedding the resulting pNJL partonic transport model in an extended multiphase transport (extended AMPT) model, we then study the elliptic flow splittings between particles and their antiparticles in relativistic heavy-ion collisions at RHIC-BES energies. We find that a weak quark vector interaction in the partonic phase is able to describe the elliptic flow splitting between protons and antiprotons in heavy-ion collisions at $\sqrt{s_{NN}}=7.7$ to 39 GeV. Knowledge on the quark vector interaction is useful for understanding the equation of state of quark matter at large baryon chemical potentials and thus the location of the critical point in the QCD phase diagram.
\end{abstract}

\maketitle

\section{INTRODUCTION}
\label{introduction}

Understanding the properties of quark-gluon plasma (QGP) and the phase diagram of the quantum chromodynamics (QCD) are among the main goals of the experiments with heavy ions at relativistic energies.  For a QGP of low baryon chemical potentials produced at ultrarelativistic heavy-ion collisions, its transition to the hadronic matter is a smooth crossover according to the lattice QCD (LQCD) calculations~\cite{Ber05,Aok06,Baz12a}. Although the LQCD has not been able to determine the properties of QGP at high baryon chemical potentials, it was shown in various effective models for QCD, such as the Nambu-Jona-Lasinio (NJL) model~\cite{Asa89,Fuk08,Car10,Bra13}, the Dyson-Schwinger approach~\cite{Xin14,Fis14}, and the functional renormalization group method~\cite{Fu19,Gao20}, that its transition to the hadronic matter could be a first-order one. Searching for the QCD critical point, at which the smooth crossover changes to a first-order transition, and determining its location in the QCD phase diagram, is currently a topic of great interest in high-energy nuclear physics, particularly in relativistic heavy-ion collisions that are carried out at RHIC-BES and will also be carried out at FAIR-CBM as well as at NICA and HIAF. On the other hand, observables measured in heavy-ion collisions at lower energies can also provide information on the properties of quark matter, especially its equation of state (EOS), at large baryon chemical potentials. Such information is helpful for understanding the properties of strange quark stars~\cite{Chu15,Liu20} and the gravitational wave from compact star mergers. We note that the QCD phase diagram and the quark matter EOS at large baryon chemical potentials are related through the underlying quark interactions.

To search for the QCD critical point from the many interesting phenomena observed in the experiments at the beam-energy scan program at RHIC, there have been great theoretical efforts based on various transport~\cite{PHSD,PHQMD,SMASH} or hydrodynamic~\cite{Hydrop,Shen} approaches. Assuming that the trajectory traversed by the produced matter in the QCD phase diagram could pass through the QCD critical region or the spinodal region, there will then be observational effects on conserved charge fluctuations~\cite{Ste09,Asa09,STAR10,STAR14,STAR20} and light cluster productions~\cite{Sun18,Sun20,Zha21} as well as other physical quantities. These studies also provide the possibility to study the quark matter EOS and the underlying quark interactions, especially at finite baryon chemical potentials. For example, from the collective flow splittings between particles and their antiparticles, one can determine the different mean-field potentials acting on quarks and antiquarks in the baryon-rich quark matter and thus their interactions. Using a QCD effective model with these constrained quark interaction, one can then construct the QCD phase diagram in the temperature and chemical potential plane, besides determining the location of the critical point.

Using an extended multiphase transport (extended AMPT) model with its partonic stage replaced by the partonic transport model based on the 3-flavor NJL model~\cite{Xu12,Xu14,Xu16}, we have previously studied the elliptic flow splittings between particles and their antiparticles in relativistic heavy-ion collisions at RHIC-BES energies.  We have found that the data measured by the STAR Collaboration~\cite{STAR13} can be explained by choosing a proper strength for the vector interaction in the NJL model. The vector interaction extracted from our study puts a significant constraint not only on the EOS of quark matter at finite baryon chemical potentials but also on the properties of the QCD phase diagram. It is, however, well known that the NJL model gives a relative low temperature for the critical point due to the lack of the effects from gluons. This drawback can be resolved by modifying the NJL model to include the contribution from the Polyakov loops~\cite{Fuk04}, which then allows for the existence of the deconfinement transition besides the chiral transition in the NJL model. In the present manuscript, we report on the results from an improved transport model study of relativistic heavy-ion collisions based on the Polyakov-looped NJL (pNJL) model. Details on the extended AMPT model as well as the NJL and the pNJL model are provided in Sec.~\ref{theory}. Section~\ref{results} gives detailed results on the beam energy dependence of heavy-ion collision dynamics and the $v_2$ splittings between particles and their antiparticles, as well as those on the QCD phase diagram. We conclude our study in Sec.~\ref{summary} and also include in the Appendices the details on the treatment of the collision integral in the partonic transport model, particularly the one based on the pNJL model.

\section{Theoretical framework}
\label{theory}

The present study is based on the framework of an extended AMPT model with its description of the partonic dynamics using the NJL transport model. The extended AMPT model has been used previously to study the splittings between the elliptic flows~\cite{Xu14,Xu16} as well as directed flows~\cite{Guo18} of particles and their antiparticles in heavy-ion collisions at RHIC-BES energies. In the present study, we further extend the NJL transport model to include the effect of Polyakov loops and make additional improvements to the extended AMPT model.

\subsection{An extended multiphase transport model}
\label{AMPT}

As in the original string-melting AMPT model~\cite{Lin05}, the momentum distribution of initial partons in the present study is obtained from constituent quarks and antiquarks of hadrons produced from elastic and inelastic scatterings of participant nucleons in the heavy-ion jet interaction generator (HIJING) model~\cite{Xnw91}. The positions of these initial partons in the transverse plane are set to be the same as those of the colliding nucleons from which they are produced. For their longitudinal positions, they are taken to be uniformly distributed within $(-lm_N/\sqrt{s_{NN}}, lm_N/\sqrt{s_{NN}})$, where $m_N=0.938$ MeV is nucleon mass, $\sqrt{s_{NN}}$ is the collision energy per nucleon pair in the center-of-mass frame, and $l=\sqrt{D^2-\rm{b}^2}$ fm is the thickness parameter with $D$ being the diameter of the colliding nucleus ($D=14$ fm for Au nucleus in the present study) and $\rm{b}$ being the impact parameter. With the initial partons gradually formed according to their formation time~\cite{Lin05}, the evolution of these partons is then described by the NJL transport model or its extension by the inclusion of Polyakov loops, where both the mean-field potentials on partons and their scatterings are properly taken into account. For the mean-field potentials, they are calculated by using the test-particle method with parallel events, while scatterings between partons are treated with the stochastic method~\cite{Xu05} using an isotropic cross section determined by the specific shear viscosity of the quark matter and including the Pauli blocking factors for fermions, as described in Appendices \ref{xsection} and \ref{stochastic}. The value of the specific shear viscosity is adjusted so that the transport model reproduces the final charged-particle elliptic flow. Details of the NJL and pNJL model will be given in the next subsection.

The partonic phase ends when the parton energy density in the central region of the system drops to about 0.8 GeV/fm$^3$. The hadronization of partons is achieved by coalescing partons close in phase space into hadrons~\cite{Wan19}, during which the baryon and electric charges as well as strangeness are conserved, although the numbers of mesons, baryons, and antibaryons can be different from the initial ones from HIJING. The evolution of the formed hadronic matter is described by a relativistic transport (ART) model~\cite{Li95}, which includes various of elastic and inelastic channels as well as decay channels for resonances, with the conserved electric charge~\cite{Xu16} and mean-field potentials included for baryons, antibaryons, pions, and kaons~\cite{Xu12}. Typically, nucleons are affected by a weakly attractive potential while antinucleons are affected by a strongly attractive potential in baryon-rich hadronic matter according to the study of subthreshold production of antiprotons~\cite{Li94}. For other baryons (antibaryons), their potentials are reduced with respect to nucleons (antinucleons) by their light quark (antiquark) fractions. For the kaon and antikaon potentials, they are weakly repulsive and strongly attractive in baryon-rich hadronic matter, respectively, according to the chiral effective field theory~\cite{Li97}.

\subsection{Nambu-Jona-Lasinio model and its Polyakov-loop extension}
\label{NJL}

We start from the Lagrangian of the 3-flavor NJL model, which is written as
\begin{eqnarray}
\mathcal{L}_{\rm NJL} &=& \bar{\psi}(i\rlap{\slash}\partial-\hat{m})\psi
+\frac{G_S}{2}\sum_{a=0}^{8}[(\bar{\psi}\lambda_a\psi)^2+(\bar{\psi}i\gamma_5\lambda_a\psi)^2]
\notag\\
&-&\frac{G_V}{2}\sum_{a=0}^{8}[(\bar{\psi}\gamma_\mu\lambda_a\psi)^2+
(\bar{\psi}\gamma_5\gamma_\mu\lambda_a\psi)^2]
\notag\\
&-&K\{\det[\bar{\psi}(1+\gamma_5)\psi]+\det[\bar{\psi}(1-\gamma_5)\psi]\}.
\end{eqnarray}
In the above, $\psi = (u, d, s)^T$ and $\hat{m} = \text{diag}(m_u, m_d, m_s)$ are the quark fields and the current quark mass matrix for $u$, $d$, and $s$ quarks, respectively; $\lambda_a$ are the Gell-Mann matrices with $\lambda_0$ = $\sqrt{2/3}I$ in the 3-flavor space; $G_S$ and $G_V$ are, respectively, the scalar and vector coupling constant; and the term with the coupling constant $K$ represents the six-point Kobayashi-Maskawa-t' Hooft interaction that breaks the axial $U(1)_A$ symmetry. In the present study, we employ the parameters $m_u = m_d = 3.6$ MeV, $m_s = 87$ MeV, $G_S\Lambda^2 = 3.6$, $K\Lambda^5 = 8.9$, and the cutoff value in the momentum integral $\Lambda = 750$ MeV/c given in Refs.~\cite{Bra13,Lut92}. The position of the critical point is sensitive to the value of $G_V$~\cite{Asa89,Fuk08,Bra13}, which can thus be used to characterize the QCD phase diagram and the quark matter EOS at large baryon chemical potentials, as detailed later. For the vector coupling $G_V$, we choose typical values of the reduced vector coupling constant $R_V=G_V/G_S=0$, 0.5 and 1.1, corresponding to the case without vector interaction, that from the Fierz transformation, and that from reproducing vector meson masses~\cite{Lut92}, respectively.

From the mean-field approximation and assuming that the vector coupling is flavor independent, the thermodynamic potential $\Omega_\textrm{NJL}$ of a quark matter at the temperature $T$ and the chemical potential $\mu_q$ for quark flavor $q$ can be expressed in terms of the in-medium Dirac mass $M_q$ and energy $E_q$ of a quark of flavor $q$ as well as the effective chemical potential $\tilde\mu_q$ and condensate $\phi_q$ of quarks of flavor $q$, i.e.,
\begin{eqnarray}\label{omeganjl}
\Omega_{\textrm{NJL}}
&=& -2N_c\sum_{q=u,d,s}\int_0^\Lambda\frac{d^3p}{(2\pi)^3}[E_q \notag\\
&+&T\ln(1+e^{-\beta(E_q-\tilde{\mu}_q)})+T\ln(1+e^{-\beta(E_q+\tilde{\mu}_q)})] \notag\\
&+&G_S(\phi_u^2+\phi_d^2+\phi_s^2)-4K\phi_u\phi_d\phi_s-\frac{1}{3}G_V\rho^2.
\end{eqnarray}
In the above, the factor $2N_c=6$ represents the spin and color degeneracy of the quark, and $\beta=1/T$ is the inverse of the temperature.

The quark condensate $\phi_q$ is given by the integral,
\begin{equation}\label{sigmaq}
\phi_q=-2N_c\int_0^{\Lambda}\frac{d^3p}{(2\pi)^3}\frac{M_q}{E_q}(1-f_q-\bar{f_q}),
\end{equation}
where $f_q$ and $\bar{f_q}$ denote, respectively, the phase-space distribution functions of quarks and antiquarks. For a static quark matter in thermal equilibrium, they are given by the Fermi-Dirac distributions,
\begin{eqnarray}\label{fermi}
f_q &=& \frac{1}{\exp[(E_q-\tilde{\mu}_q)/T]+1},\label{fqnjl}\notag\\
\bar{f_q} &=& \frac{1}{\exp[(E_q+\tilde{\mu}_q)/T]+1},\label{fqbnjl}
\end{eqnarray}
where the effective chemical potential $\tilde{\mu}_q$ is related to the chemical potential $\mu_q$ through the relation
\begin{equation}\label{muq}
\tilde{\mu}_q = \mu_q-\frac{2}{3}G_V\rho. \\
\end{equation}
In the above, $\rho=\rho_u+\rho_d+\rho_s$ is the total net quark number density, and for a single quark flavor $q$ the net quark number density $\rho_q$ can be calculated from
\begin{equation}\label{rhoqnjl}
\rho_q=2N_c\int^{\Lambda}_0\frac{d^3p}{(2\pi)^3}(f_q-\bar{f_q}).
\end{equation}

The single-quark energy $E_q$ in Eqs.~(\ref{omeganjl}), (\ref{sigmaq}), and (\ref{fermi}) is given by  $E_q=\sqrt{M_q^2+{\vec p}^2} $ with $\vec{p} = \vec{p}^{\ast} \mp \frac{2}{3}G_V{\vec j}$ being the momentum of the parton, where $\vec{p}^{\ast}$ is the canonical momentum and $\vec{j}=\vec{j}_u+\vec{j}_d+\vec{j}_s$ is  the net quark current with that for each quark flavor $q$ given by
\begin{equation}\label{currentnjl}
\vec{j}_q = 2N_c\int_0^\Lambda\frac{d^3p}{(2\pi)^3} \frac{\vec{p}}{E_q}(f_q-{\bar f}_q).
\end{equation}

The in-medium Dirac mass $M_q$ in Eq.~(\ref{sigmaq}) is related to the quark condensate through the relations
\begin{eqnarray}
M_{u} &=& m_{u}-2G_S\phi_{u}+2K\phi_{d}\phi_{s},\label{mass1}\notag\\
M_{d} &=& m_{d}-2G_S\phi_{d}+2K\phi_{s}\phi_{u}, \label{mass2}\notag\\
M_{s} &=& m_{s}-2G_S\phi_{s}+2K\phi_{u}\phi_{d}. \label{mass3}
\end{eqnarray}
Since the quark condensates and the Dirac masses are related to each other, Eqs.~(\ref{sigmaq}) and (\ref{mass3}) need to be solved self-consistently through the iteration method. We note that the quark condensate or the Dirac mass is the order parameter for the chiral phase transition.

Equation~(\ref{omeganjl}) can be generalized to obtain the thermodynamic potential $\Omega_\textrm{pNJL}$ of the 3-flavor pNJL model at finite temperature and quark chemical potential, and it is
\begin{eqnarray}\label{omegapnjl}
\Omega_{\textrm{pNJL}} &=&\mathcal{U}(\Phi,\bar{\Phi},T)-2N_c\sum_{q=u,d,s}\int_0^{\Lambda}\frac{d^3p}{(2\pi)^3}E_q
\notag\\
&-&2T\sum_{q=u,d,s}\int\frac{d^3p}{(2\pi)^3}[\ln(1+e^{-3\beta(E_q-\tilde{\mu}_q)}
\notag\\
&+&3\Phi e^{-\beta(E_q-\tilde{\mu}_q)}
+3\bar{\Phi}e^{-2\beta(E_q-\tilde{\mu}_q)})
\notag\\
&+&\ln(1+e^{-3\beta(E_q+\tilde{\mu}_q)}
+3\bar{\Phi} e^{-\beta(E_q+\tilde{\mu}_q)}
\notag\\
&+&3\Phi e^{-2\beta(E_q+\tilde{\mu}_q)})]
+G_S(\phi_u^2+\phi_d^2+\phi_s^2)
\notag\\
&-&4K\phi_u\phi_d\phi_s-\frac{1}{3}G_V\rho^2.
\end{eqnarray}
In the above, $\Phi$ ($\bar{\Phi}$) is the contribution from the Polyakov loop, which is related to the excess free energy due to a static quark (anti-quark) in a hot gluon medium~\cite{Fuk11} and thus serves as an order parameter for the deconfinement phase transition. The form of the temperature-dependent effective potential
$\mathcal{U}(\Phi, \bar{\Phi}, T)$ as a function of the Polyakov loops $\Phi$ and $\bar{\Phi}$ in Eq.~(\ref{omegapnjl}) is taken from Ref.~\cite{Fuk08}, i.e.,
\begin{eqnarray}
\mathcal{U}(\Phi,\bar{\Phi},T) &=& -b
T\{54e^{-a/T}\Phi\bar{\Phi} +\ln[1-6\Phi\bar{\Phi}
\notag\\
&-&3(\Phi\bar{\Phi})^2+4(\Phi^3+\bar{\Phi}^3)]\}.
\end{eqnarray}
The parameters $a=664$ MeV and $b=0.028\Lambda^3$ are determined by the condition
that the first-order phase transition in a pure gluon matter takes place at $T = 270$ MeV~\cite{Fuk08}, and the simultaneous crossover of the chiral restoration and the deconfinement phase transition occurs around $T \approx 212$ MeV.

Minimizing the thermodynamic potential $\Omega_\textrm{pNJL}$ with respect to the quark condensates, i.e.,
\begin{eqnarray}
\frac{\partial\Omega_{\textrm{pNJL}}}{\partial\phi_u}
=\frac{\partial\Omega_{\textrm{pNJL}}}{\partial\phi_d}
=\frac{\partial\Omega_{\textrm{pNJL}}}{\partial\phi_s}
=0,
\end{eqnarray}
leads to the similar expression for $\phi_u$, $\phi_d$, and $\phi_s$
as in Eq.~(\ref{sigmaq}) except the absence of the cutoff in the momentum integral and the replacement of the quark and antiquark phase-space distribution functions $f_q$ and ${\bar f}_q$ by $F_q$ and ${\bar F}_q$, respectively, given by
\begin{eqnarray}\label{Fq}
F_q &=& \frac{1+2\bar\Phi\xi_q+\Phi\xi_q^2}{1+ 3\bar\Phi\xi_q+3\Phi\xi_q^2+\xi_q^3},\notag\\
\bar{F_q}&=& \frac{1+2\Phi{\xi^\prime_q}+\bar\Phi{\xi^\prime_q}^2}{1+3\Phi{\xi^\prime_q}+3\bar\Phi{\xi^\prime_q}^2+{\xi^\prime_q}^3},
\end{eqnarray}
with $\xi_q =e^{(E_q-\tilde{\mu}_q)/T}$ and $\xi_q^\prime =e^{(E_q+\tilde{\mu}_q)/T}$. Similarly, the net quark density and current given by Eq.~(\ref{rhoqnjl}) and Eq.~(\ref{currentnjl}), respectively, in the NJL model are modified by replacing $f_q$ and ${\bar f}_q$ with $F_q$ and ${\bar F}_q$, respectively, and without the cutoff in the momentum integrals. For the values of the Polyakov loops $\Phi$ and $\bar\Phi$ in Eq.~(\ref{Fq}), they are determined by minimizing the thermodynamic potential $\Omega_{\rm pNJL}$ with respect to $\Phi$ and $\bar\Phi$, i.e., $\frac{\partial\Omega_{\textrm{pNJL}}}{\partial\Phi}=\frac{\partial\Omega_{\textrm{pNJL}}}{\partial\bar{\Phi}}=0$.

Starting from the thermodynamic potential, the energy density of the system can be obtained from the thermodynamical relation
\begin{equation}
\varepsilon=\Omega+\beta\frac{\partial}{\partial\beta}\Omega+\sum_{q=u,d,s}\mu_q\rho_q.
\end{equation}
For the NJL model, the energy density is then
\begin{eqnarray}\label{epsilon}
\varepsilon_{\textrm{NJL}}&=&-2N_c\sum_{q=u,d,s}\int_0^{\Lambda}\frac{d^3p}{(2\pi)^3}
E_q(1-f_q-\bar{f_q})
\notag\\
&+&G_S(\phi_u^2+\phi_d^2+\phi_s^2)-4K\phi_u\phi_d\phi_s\notag\\
&+&\frac{1}{3}G_V\rho^2 - \varepsilon_0,
\end{eqnarray}
where the last term $\varepsilon_0$ is from the quark condensate contribution to the energy density in vacuum and is needed to ensure $\varepsilon_{\textrm{NJL}} = 0$ in vacuum.

Similarly, the energy density from the pNJL model can be expressed as
\begin{eqnarray}\label{epsilon_pNJL}
\varepsilon_{\textrm{pNJL}}&=&
54abe^{-a/T}\Phi\bar{\Phi}-2N_c\sum_{q=u,d,s}\int_0^{\Lambda}\frac{d^3p}{(2\pi)^3}
E_q
\notag\\
&+&2N_c\sum_{q=u,d,s}\int\frac{d^3p}{(2\pi)^3}
E_q(F_q+\bar{F_q})\notag\\
&+&G_S(\phi_u^2+\phi_d^2+\phi_s^2)-4K\phi_u\phi_d\phi_s\notag\\
&+&\frac{1}{3}G_V\rho^2-\varepsilon_0.
\end{eqnarray}

For both the NJL and the pNJL model, the pressure $P$ can be calculated from the thermodynamic potential according to
\begin{equation}
P = -\Omega + \Omega_0,
\end{equation}
where $\Omega_0=\varepsilon_0$ is the thermodynamic potential in vacuum to ensure that the pressure is zero in vacuum.

\begin{figure}[ht]
	\includegraphics[scale=0.5]{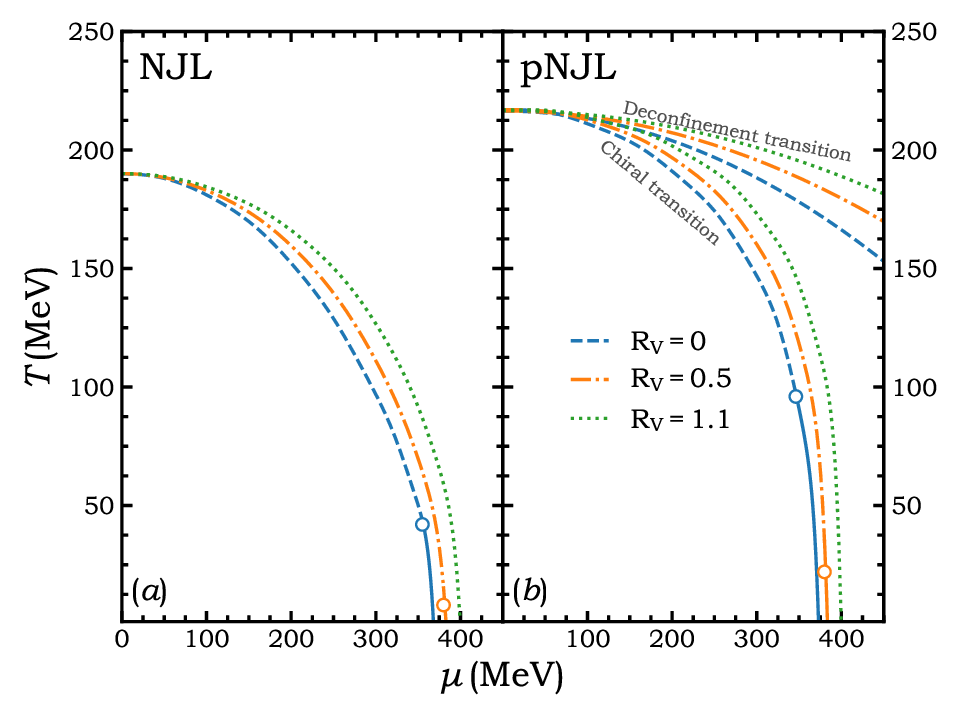}\\
    \includegraphics[scale=0.5]{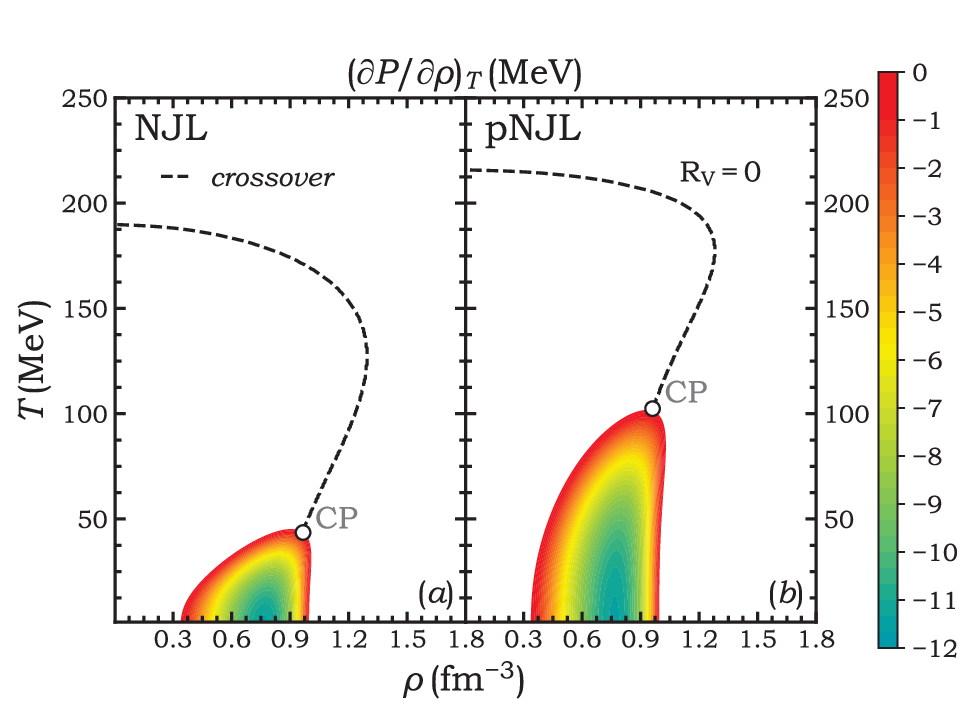}
	\caption{(Color online) Upper: Phase diagram in the $\mu-T$ plane from the NJL (left) and the pNJL (right) model with $\mu=\mu_u=\mu_d=\mu_s$ for different values of the reduced vector coupling constant $R_V=G_V/G_S$. Lower: Phase diagram as well as the $(\partial P/\partial \rho)_T<0$ region in the $\rho-T$ plane from the NJL (left) and the pNJL (right) model with $\mu=\mu_u=\mu_d=\mu_s$ without vector coupling. `CP' in the lower panels represents the critical point.}\label{fig1}
\end{figure}

The chiral phase transition boundary in the quark chemical potential and temperature ($\mu-T$) plane for different values of the reduced vector coupling constant $R_V$ are displayed in the upper panels of Fig.~\ref{fig1} for both the NJL and the pNJL model. At small $\mu$ and low $T$, the system is in the chiral symmetry broken phase with large in-medium light quark masses $M_q$. At large $\mu$ or high $T$, the system is in the chiral symmetry restored phase with small in-medium light quark masses $M_q \sim m_q$. The dashed and the solid lines in Fig.~\ref{fig1} denote, respectively, the boundary between the chiral symmetry broken and restored phases that are connected by a crossover and a first-order transition. The point where the dashed and the solid line meets is the critical point in the QCD phase diagram. The corresponding critical temperature is seen to decrease with increasing value of $R_V$, as a result of the increasing stiffness of the EOS of baryon-rich quark matter. To better understand the relation between the EOS and the QCD critical point, we also plot in the lower panels of Fig.~\ref{fig1} the phase diagram in the $\rho-T$ plane without the vector coupling for both NJL and pNJL models. Also shown in these panels is the mechanical instability or the spinodal region with $(\partial P/\partial \rho)_T<0$, where small density fluctuations are expected to grow exponentially in time. The spinodal region or the hadron-quark mixed phase shrinks with increasing temperature and disappears when the temperature is above the critical temperature. With the increasing value of $R_V$, the spinodal region also shrinks with the critical point moving to a lower temperature. Due to the different phase-space distribution functions in the pNJL model [Eq.~(\ref{Fq})] and the NJL model [Eq.~(\ref{fqnjl})], the effective temperature of the quark matter is higher in the pNJL model than in the NJL model of the same vector coupling. This leads to a higher critical temperature in the pNJL model, that is more comparable with those predicted by the Dyson-Schwinger approach~\cite{Xin14,Fis14} and the functional renormalization group method~\cite{Fu19,Gao20}.

\subsection{NJL and pNJL parton transport models}

For the NJL model, the single-particle Hamiltonian for a parton of flavor $q(q=u,d,s)$ is given by
\begin{eqnarray}\label{eq2}
H_q &=& \sqrt{M_q^2+\vec{p}^2}\pm \frac{2}{3}G_V\rho.
\end{eqnarray}
This then leads to the following transport equation for the quark and antiquark phase-space distribution functions $f_q({\vec r},{\vec p})$~\cite{Ber88}:
\begin{eqnarray}\label{transport}
\frac{\partial f_q}{\partial t}+\frac{{\vec p}_q}{E_q}\cdot{\vec\nabla}f_q-{\vec\nabla}H_q\cdot{\vec\nabla}_pf_q=I_{\rm coll},
\end{eqnarray}
where the collision integral $I_{\rm coll}$ describes the effect of quark scatterings on the quark phase-space distribution functions, and it depends on the scattering cross section and the Pauli-blocking factors in the final state of a scattering. Solving Eq.~(\ref{transport}) by the test-particle method using an ensemble of parallel events~\cite{Ber88,Won82}, we obtain the following canonical equations of motion for partons between their scatterings:
\begin{eqnarray}
\frac{d\vec{r}}{dt} &=& \frac{\vec{p}}{E_q}, \label{eom1} \notag\\
\frac{d\vec{p}}{dt} &=& -\frac{M_q}{E_q} \vec\nabla M_q \pm \left(\vec{E}_q+\frac{\vec{p}}{E_q} \times \vec{B}_q\right), \label{eom2}
\end{eqnarray}
where
\begin{eqnarray}
\vec{E}_q = - \frac{2}{3} G_V\left(\vec\nabla \rho + \frac{\partial \vec{j}}{\partial t}\right)
\end{eqnarray}
is the effective electric field, and
\begin{equation}
\vec{B}_q = \frac{2}{3} G_V \vec\nabla \times \vec{j}
\end{equation}
is the effective magnetic field.

For the parton scattering cross section in the collision integral of the transport equation, it is obtained from the specific shear viscosity of the partonic matter as described in Appendix~\ref{xsection}. As to the treatment of parton scattering with the Pauli blocking factor, we use the stochastic method~\cite{Xu05} based on a full ensemble of $N_{TP}$ parallel events as detailed in Appendix~\ref{stochastic}.

The quark and antiquark phase-space distribution functions $f_q$ and $\bar{f_q}$ in the transport model simulations are calculated by summing the particles from $N_{TP}$ parallel events in the local rest frame of the cells around $(\vec{r},\vec{p})$ in the phase space, i.e.,
\begin{eqnarray}\label{phase-space}
f_q(\vec{r},\vec{p}) &\sim& \frac{1}{N_{TP}}\sum_{i\in q} \delta(\vec{r}-\vec{r}_i)\delta(\vec{p}-\vec{p}_i),\notag\\
\bar{f_q}(\vec{r},\vec{p}) &\sim& \frac{1}{N_{TP}}\sum_{i \in \bar{q}} \delta(\vec{r}-\vec{r}_i)\delta(\vec{p}-\vec{p}_i).
\end{eqnarray}

\begin{figure}[ht]
	\includegraphics[scale=0.6]{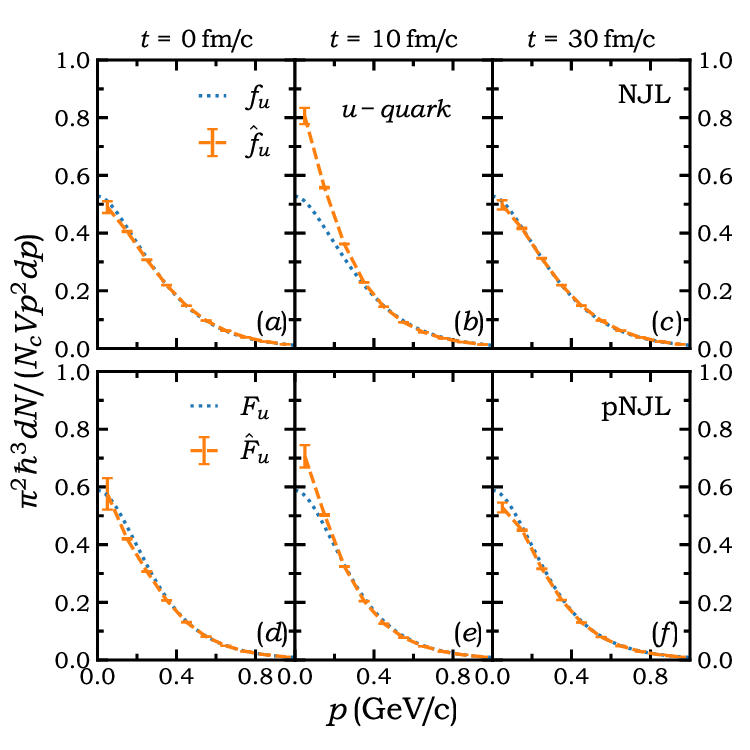}
	\caption{(Color online) Momentum distribution of $u$ quark at different times for quark matter in a box with periodic boundary conditions. The initial distribution at $t=0$ is that of the thermally equilibrated distribution given by Eq.~(\ref{fqnjl}) for the NJL model and Eq.~(\ref{Fq}) for the pNJL model (dotted lines). At $t=10$ fm/$c$, it reaches the Boltzmann distribution due to the neglect of Pauli blockings in the scatterings. After turning on Pauli blockings at this time, the distribution changes back to the initial distribution as seen at $t=30$ fm/$c$. See text for details.} \label{fig2}
\end{figure}

The NJL-based transport equation given in Eq.~(\ref{transport}) and the resulting equations of motion for quarks and antiquarks can be generalized to that based on the pNJL model by replacing the quark and antiquark phase-space distribution functions $f_q$ and $\bar{f_q}$ with $F_q$ and ${\bar F}_q$, respectively. A modified treatment of the Pauli blocking factor in partonic scattering is, however, required to ensure the approach of $F_q$ and ${\bar F}_q$ to the thermally equilibrated values given in Eq.~(\ref{Fq}) as described in Appendix~\ref{collisions}. To illustrate the validity of this method, we consider the evolution of a partonic matter of uniform spatial distribution of density $\rho=1.5$ fm$^{-3}$ in a box with periodic boundary conditions. Starting from the Fermi-Dirac momentum distributions of Eq.~(\ref{fqnjl}) for the NJL model or the momentum distributions of Eq.~(\ref{Fq}) for the pNJL model at temperature $T=200$ MeV, the system evolves in time by solving the NJL  or pNJL transport model without the Pauli blocking factors in parton scatterings for the first 10 fm/$c$. As shown in Fig.~\ref{fig2} for $u$ quarks as an example, the momentum distribution evolves into a Boltzmann distribution shown by the dashed line, which is obviously different from the initial momentum distribution shown by the dotted line. Turning on the Pauli blocking factors at $t=10$ fm/$c$, it is seen that the momentum distribution gradually changes back to the initial ones, demonstrating the correctness of our treatment of the collision integral in the NJL and pNJL models.

\section{Results and discussions}
\label{results}

Based on the extended AMPT model with the pNJL transport model to describe the evolution of the partonic phase, we now study the elliptic flow ($v_2$) splittings between particles and their antiparticles in relativistic heavy-ion collisions at RHIC-BES energies. Although there have been a number of possible explanations for the observed $v_2$ splittings in the STAR experiments~\cite{Dun11,Gre12,Iva13,Ste12,Hat15,Sun15,Bur11}, our studies using different mean-field potentials for particles and their antiparticles in the baryon-rich matter are the only one that is based on a comprehensive transport model~\cite{Xu12,Xu14,Xu16}. Taking the case in the partonic phase as an example, antiquarks are generally affected by a more attractive mean-field potential than quarks in the almond-shaped baryon-rich partonic matter. Thus they tend to stay longer in the produced matter, compared with quarks which move along the short axis of this matter. Because of the difference in their mean-field potentials, quarks then generally have a larger $v_2$ than antiquarks. Such $v_2$ splittings are expected to be preserved after hadronization with a proper coalescence algorithm~\cite{Wan19} and after hadronic evolution with different hadronic mean-field potentials for baryons and antibaryons as well as for $K^+$ and $K^-$ and other particle and antiparticle pairs~\cite{Xu12}.

\subsection{Time evolution of produced matter in relativistic heavy-ion collisions}

\begin{figure}[h]
	\includegraphics[scale=0.55]{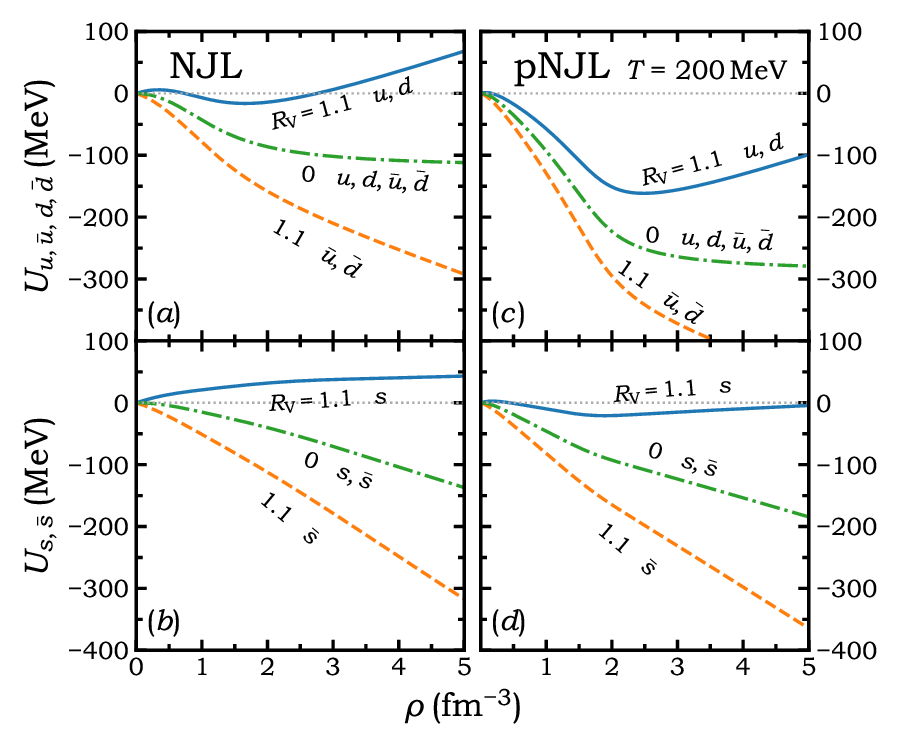}
	\caption{(Color online) Mean-field potentials of quarks and antiquarks of different flavors from non-relativistic reduction as a function of net quark density with $\mu_u=\mu_d=\mu_s$ and at the temperature $T=200$ MeV for the NJL (left) and the pNJL (right) model. } \label{fig3}
\end{figure}

Besides scatterings, the evolution of a partonic matter is also affected by the mean-field potentials of quarks and antiquarks. In the non-relativistic approximation, they can be expressed as
\begin{eqnarray}
U_{u(\bar{u})} = &-&2G_S\phi_{u}+2K\phi_{d}\phi_{s}-U_u^0 \pm \frac{2}{3} G_V \rho, \notag\\
U_{d(\bar{d})} = &-&2G_S\phi_{d}+2K\phi_{u}\phi_{s}-U_d^0 \pm \frac{2}{3} G_V \rho, \notag\\
U_{s(\bar{s})} = &-&2G_S\phi_{s}+2K\phi_{u}\phi_{d}-U_s^0 \pm \frac{2}{3} G_V \rho,
\end{eqnarray}
where the term $U_q^0$ ($q=u,d,s$) is introduced to ensure that the mean-field potential vanishes in vacuum, and the upper (lower) signs in these equations are for quarks (antiquarks). Their density dependence in a quark matter at temperature $T=200$ MeV is shown in Fig.~\ref{fig3}. It is seen that the scalar potentials shown by dash-dotted lines are attractive, as a result of the smaller effective quark masses in the medium than in vacuum. Because the pNJL model has a different thermal distribution from that in the NJL model due to the contribution from Polyakov loops, quark condensates in the pNJL model are less negative than in the NJL model at same temperature and density, leading to a more attractive scalar potential as seen in Fig.~\ref{fig3}. The mean-field potentials of quarks and antiquarks become very different when the vector coupling constant has a finite value, and the difference increases with increasing net quark density. Since the mean-field potential affects the elliptic flow $v_2$ of particles, measuring the $v_2$ splitting in heavy-ion collisions allows us to study the mean-field potential difference between quarks and antiquarks, and thus their vector interactions, which can affect significantly the EOS of quark matter at large chemical potentials and the QCD phase diagram in the NJL or pNJL model. The different mean-field potentials between $s$ quarks and anti $s$ quarks are qualitatively similar to those between light quarks and their antiquarks.

\begin{figure}[h]
	\includegraphics[scale=0.5]{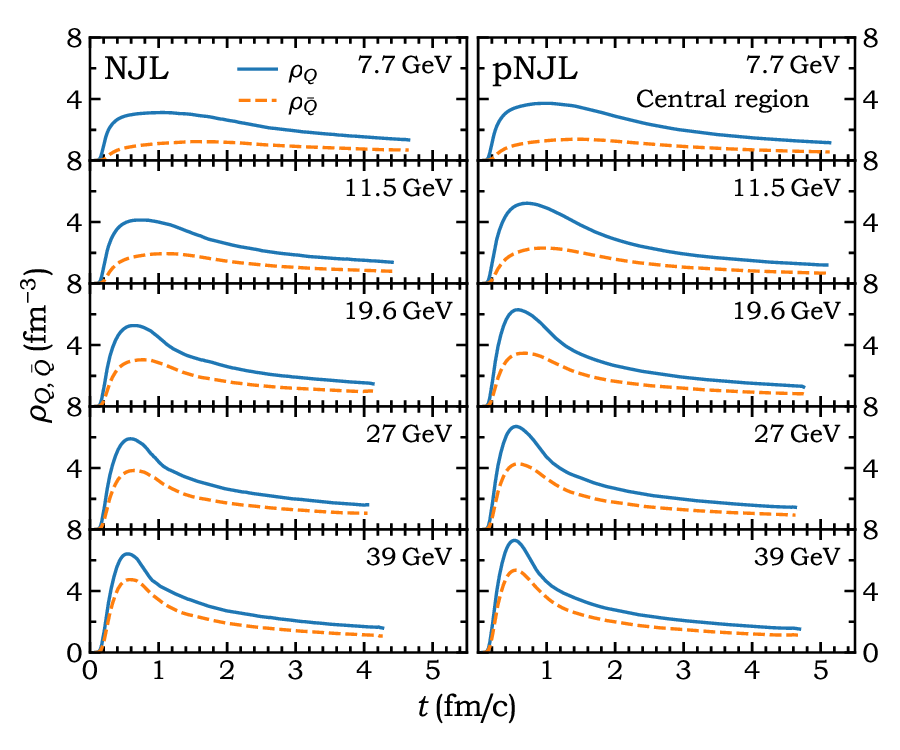}
	\caption{(Color online) Average central number densities of quarks ($\rho_Q$) and antiquarks ($\rho_{\bar{Q}}$) as a function of time in minibias ($0-80\%$) Au+Au collisions at different collision energies for both NJL (left) and pNJL (right) models. } \label{fig4}
\end{figure}

The time evolutions of average densities of quarks ($\rho_Q$) and antiquarks ($\rho_{\bar{Q}}$) in the central region of minibias Au+Au collisions at different collision energies are displayed in Fig.~\ref{fig4}, with $\rho_Q-\rho_{\bar{Q}}=\rho$ being the net quark number density. These results are obtained with a parton scattering cross section that corresponds to the specific shear viscosity of 0.14 and 0.06 in the NJL and the pNJL transport model, respectively, in order to reproduce measured $v_2$ of charged hadrons as discussed later. It is seen that partons are generally formed after 0.2 fm/$c$. With the gradual formation of partons, the central density peaks at around 0.5 fm/$c$ at higher collision energies, while at the collision energy of $\sqrt{s_{NN}}=7.7$ GeV the duration of the peak density lasts for a longer time due to the finite thickness of initial collision zone mentioned in Sec.~\ref{AMPT}. The parton phase ends around $4-5$ fm/$c$ when the central energy density drops to about 0.8 GeV/fm$^3$. In peripheral collisions, since the maximum energy density cannot reach 0.8 GeV/fm$^3$, initial partons from the AMPT model are immediately converted back to hadrons without undergoing any scatterings. The average lifetime of the partonic phase is seen to generally decrease with increasing collision energy, as a result of the stronger expansion in high-energy collisions and the quick drop of the energy density. The densities are seen to be higher in the pNJL model than in the NJL model, where the parton densities are calculated by counting those with their momenta lower than $\Lambda$ [see Eq.~(\ref{rhoqnjl})], while the lifetime of the partonic phase is slightly longer in the pNJL model than in the NJL model due to the higher energy density from the contribution of Polyakov loops.

\begin{figure}[ht]
	\includegraphics[scale=0.5]{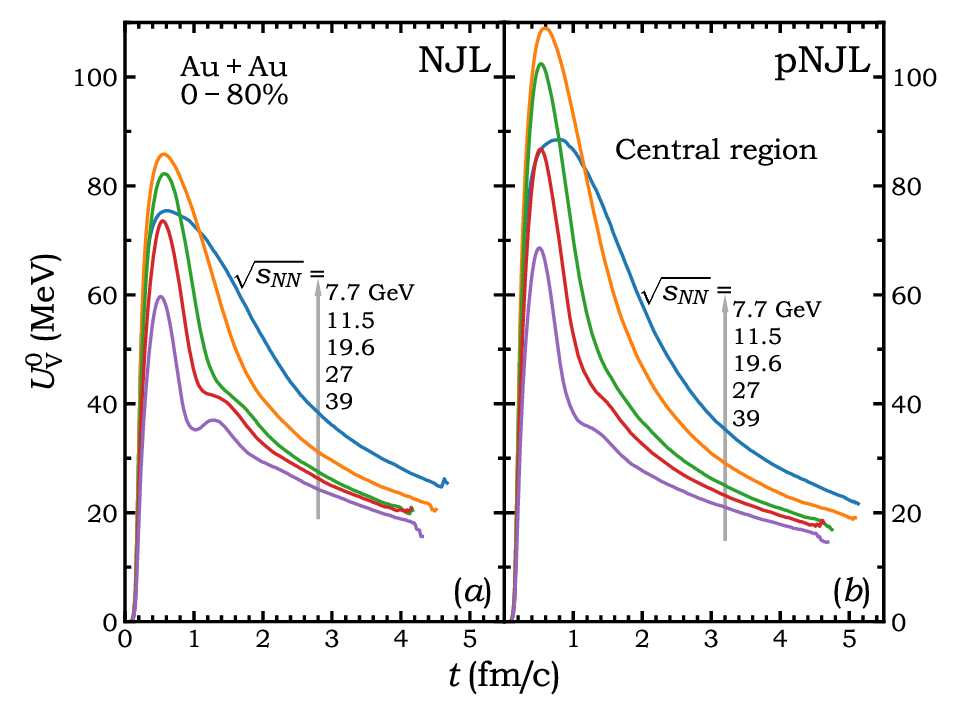}
	\caption{(Color online) Time components of the vector potential $U_V^0=\frac{2}{3} G_V \rho$ in the central region of heavy-ion collisions as a function of time at different collision energies for both NJL (left) and pNJL (right) models. } \label{fig5}
\end{figure}

The time component of the vector potential $U_V^0=\frac{2}{3} G_V \rho$ is the driving force for the $v_2$ splitting between quarks and antiquarks because the space component of the vector potential, which is proportional to the net quark current, takes time to develop and becomes important only at later stages when $v_2$ has mostly been developed. It is thus of importance to know how $U_V^0$ evolves with time, and this is displayed in Fig.~\ref{fig5} for the central region of the collision system. Although the $U_V^0$ has the longest duration at $\sqrt{s_{NN}}=7.7$ GeV, it is interesting to see that its peak value depends non-monotonically on the collision energy. The largest peak value of $U_V^0$ is reached at about $\sqrt{s_{NN}}=11.5$ GeV, due to the balance between the smaller baryon chemical potential at higher collision energies and a thicker quark matter in the longitudinal direction formed at lower collision energies. This has observational consequences on the collision energy dependence of $v_2$ splitting as shown later. The net quark density $\rho$ on which the vector potential depends is given by the difference between the quark and antiquark density shown in Fig.~\ref{fig4}, where in the NJL model only partons with their momenta lower than $\Lambda$ are counted. Since antiquarks have a much softer momentum spectrum than quarks due to their attractive mean-field potentials, the difference between the numbers of quarks and antiquarks is much larger for partons with high momenta than for those with low momenta. Thus, excluding partons with momenta higher than $\Lambda$ in the NJL model generally results in a vector potential that is weaker than that in the pNJL model in which partons of all momenta are included in calculating the net quark density, as seen in Fig.~\ref{fig4}.

\begin{figure}[h]
	\includegraphics[scale=0.5]{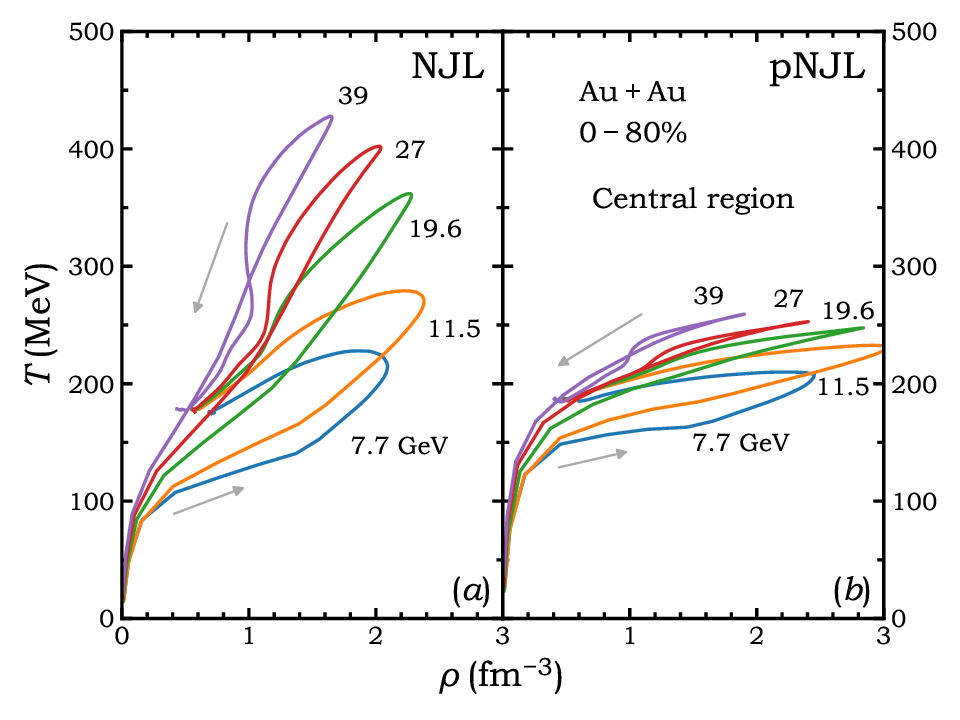}	\\
\includegraphics[scale=0.5]{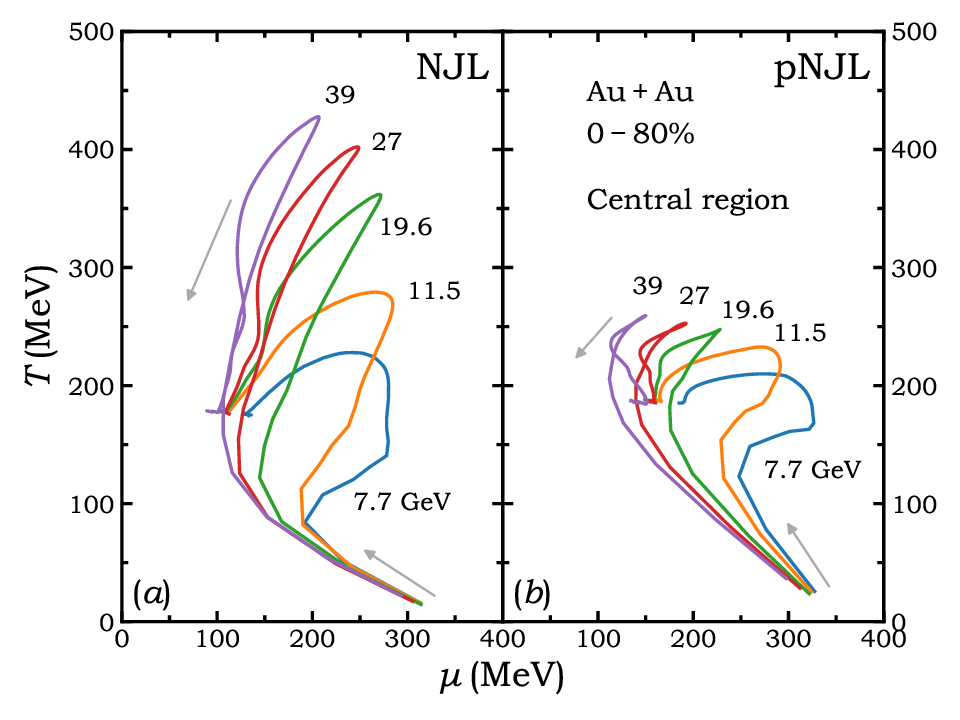}
	\caption{(Color online) Evolution trajectories of the partonic matter in the plane of temperature ($T$) versus density ($\rho$) (upper) as well as those in the plane of temperature ($T$) versus chemical potential ($\mu$) (lower) in minibias ($0-80\%$) collisions at different collision energies for both NJL (left) and pNJL (right) models. } \label{fig6}
\end{figure}

It is of great interest to know the evolution trajectory of the partonic matter formed in heavy-ion collisions in the QCD phase diagram.  Assuming that the central region of the produced matter is in thermal and chemical equilibrium, the temperature $T$ and the quark chemical potential $\mu$ can be determined from the net quark density $\rho$ and the energy density $\varepsilon$ of quarks and antiquarks by using quark and antiquark momentum distributions given by Eq.~(\ref{fermi}) for the NJL model and Eq.~(\ref{Fq}) for the pNJL model. The resulting trajectories in the ($T-\rho$) plane and ($T-\mu$) plane are displayed in Fig.~\ref{fig6} for collisions at various energies. It is seen that from the AMPT initial conditions, all trajectories follow a counterclockwise path in the phase diagram. As expected, higher temperatures are generally reached at higher collision energies. Also, the trajectories from the NJL transport model go through a higher temperature compared with that from the pNJL transport model, as a result of the different forms for the momentum distribution [Eqs.~(\ref{fqnjl}) and (\ref{Fq})] and thus different 'effective' temperatures. On the other hand, the partonic phase has a higher net quark density $\rho$ from the pNJL transport model than from the NJL model, as already shown in Figs.~\ref{fig4} and \ref{fig5}. As seen from the phase diagram shown in Fig.~\ref{fig1}, none of the trajectories passes through the spinodal region or the critical point for collision energies considered here.

\begin{figure}[h]
	\includegraphics[scale=0.5]{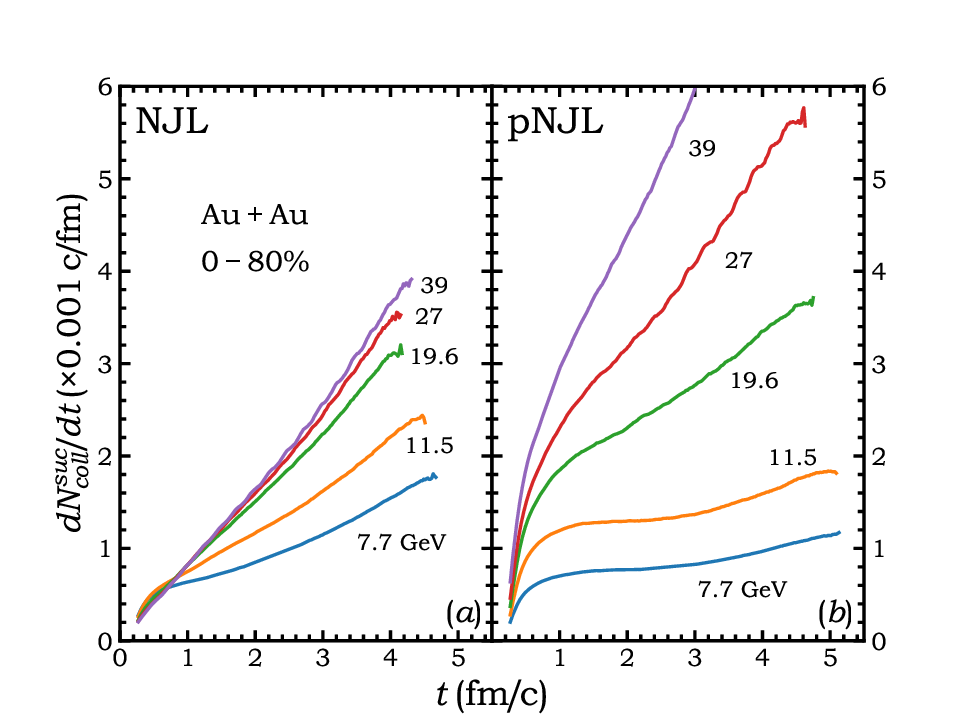}\\
    \includegraphics[scale=0.5]{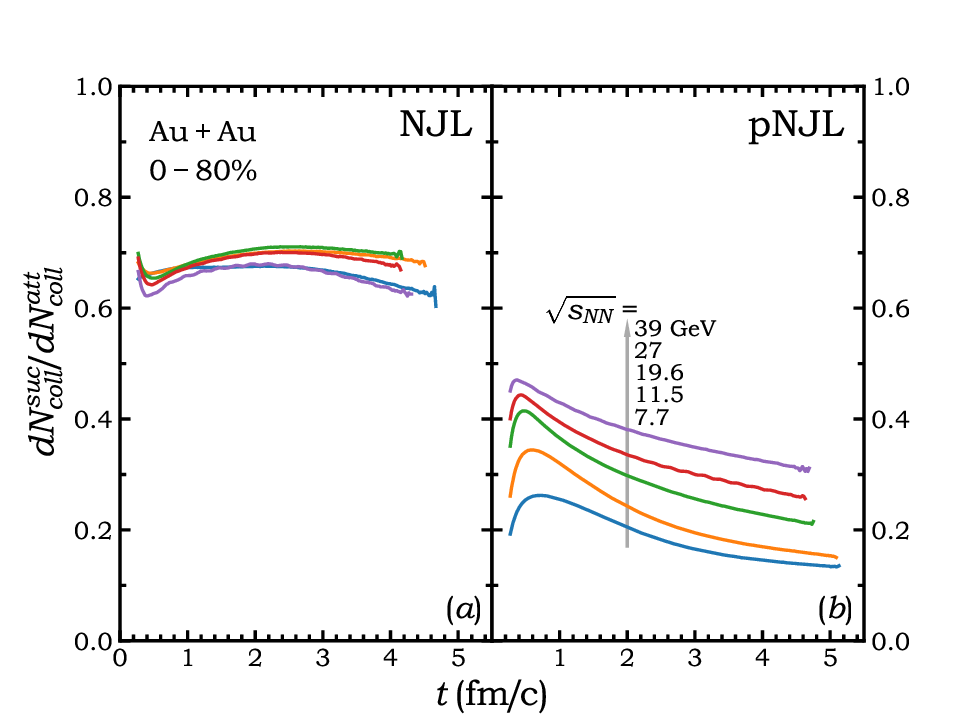}
	\caption{(Color online) Successful collision rate (upper) and ratios of successful to attempted collision rates (lower) as a function of time in the partonic phase in heavy-ion collisions at different collision energies in both NJL (left) and pNJL (right) models. } \label{fig7}
\end{figure}

As mentioned above, Pauli blockings are incorporated in the NJL transport model and also in the pNJL transport model using a modified treatment in order to ensure that correct thermal distributions are reached when the partonic matter is in thermal equilibrium. Due to the more attractive scalar potential in the pNJL model as shown in the Fig.~\ref{fig3}, a smaller specific shear viscosity $\eta/s$ and thus a larger partonic scattering cross section is used in this model than in the NJL model to obtain the same charged-particle $v_2$ in these two models. To better understand parton scatterings in the partonic phase of heavy-ion collisions, we show in Fig.~\ref{fig7} the rate of successful collisions that are not Pauli blocked as well as the ratio of successful to attempted collision rates. As expected, higher successful collision rates are observed at higher collision energies. Also, the successful collision rate generally increases with time due to the constant $\eta/s$ used in the model calculation, which results in a larger parton scattering cross section when the expanding partonic matter is at lower temperatures and/or densities (see Fig.~\ref{fig15} in Appendix~\ref{xsection}). Although a smaller $\eta/s$ is used in the pNJL transport model, the ratio of successful to attempted collision rates is much smaller than that in the NJL transport model using a larger $\eta/s$. This is again due to the different phase-space distributions and treatments of parton scatterings. Figure~\ref{fig7} further shows that while the Pauli blocking effect in the NJL model does not depend much on the collision energy, parton scatterings in the pNJL model are less blocked at higher collision energies.

\begin{figure}[h]
	\includegraphics[scale=0.5]{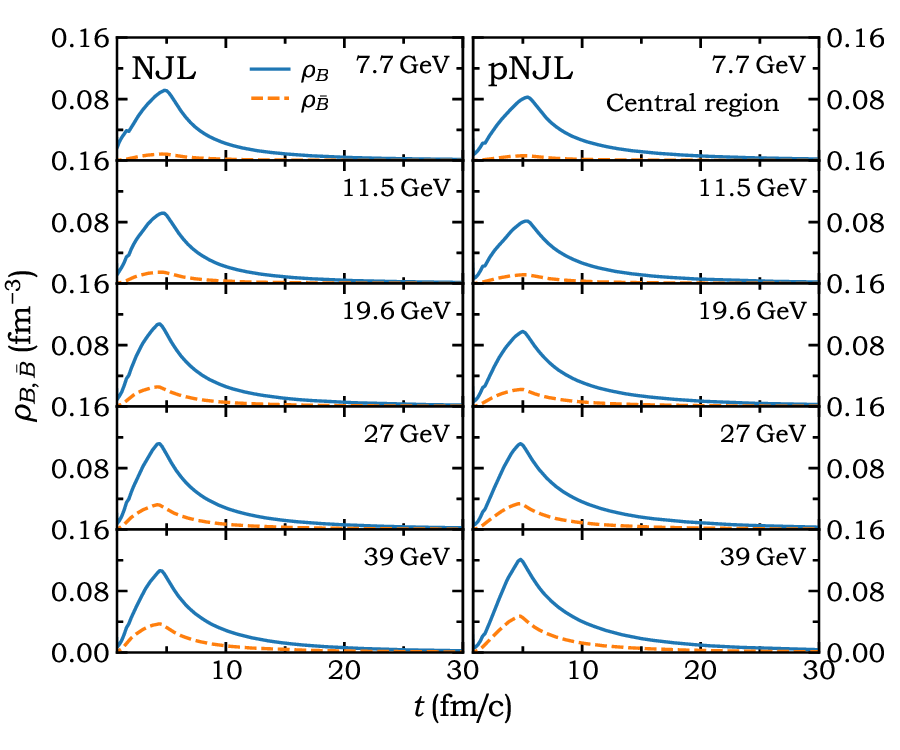}
	\caption{(Color online) Average central baryon ($\rho_B$) and antibaryon ($\rho_{\bar{B}}$) number densities in the hadronic phase as a function of time in minibias ($0-80\%$) Au+Au collisions at different collision energies from the extended AMPT model with the partonic phase described by the NJL (left) or the pNJL (right) model. } \label{fig8}
\end{figure}

For a complete description of the evolution of the matter produced in heavy-ion collisions, we show in Fig.~\ref{fig8} the time evolution of the average central baryon and antibaryon densities during the hadronic phase of a heavy-ion collision. Converting freeze-out partons via an improved hadonization scheme base on the phase-space coalescence model~\cite{Wan19}, hadrons are produced after about 5 fm/c after the production of initial partons in a heavy-ion collision. The maximum baryon density of about $0.1$ fm$^{-3}$ is reached soon after hadronization, and this value is approximately independent of the collision energy. For the maximum antibaryon density, it is reached at about the same time, but its magnitude increases with the increasing collision energy. The pNJL model generally shows similar baryon and antibaryon density evolutions compared with the NJL model. In such baryon-rich hadronic matter, antiparticles are expected to have more attractive mean-field potentials than particles~\cite{Xu12}. As a result, the $v_2$ splittings between quarks and antiquarks developed during the partonic phase can be further amplified in those between hadrons and their antiparticles.

\subsection{Description of physical observables}

\begin{figure}[h]
\includegraphics[scale=0.5]{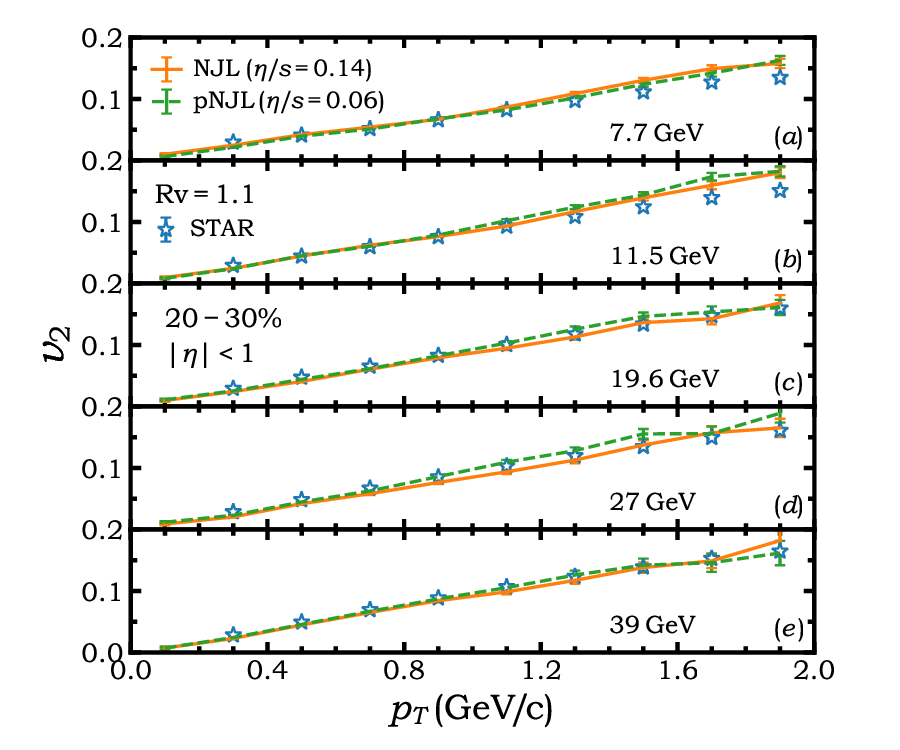}	
	\caption{(Color online) Transverse momentum dependence of mid-pseudorapidity charged-particle elliptic flow in midcentral ($20-30\%$) Au+Au collisions at different collision energies from the extended AMPT model with the partonic phase described by the NJL or the pNJL model. Experimental data are from Ref.~\cite{STAR12}. } \label{fig9}
\end{figure}

Having the evolution of produced matter in heavy-ion collisions described in the above subsection using the NJL or pNJL transport model, we show in this subsection that both models can describe satisfactorily many measured observables in heavy-ion collisions, particularly those sensitive to the bulk properties of the produced matter. Figure~\ref{fig9} shows the comparison of the transverse momentum ($p_T$) dependence of charged-particle $v_2$ at different collision energies with corresponding experimental data from the STAR Collaboration~\cite{STAR12}. By using a constant specific shear viscosity of $\eta/s=0.14$ for the NJL transport model and 0.06 for the pNJL transport model, the $p_T$ dependence of $v_2$ in midcentral ($20-30\%$) Au+Au collisions at RHIC-BES energies are reasonably well reproduced. The smaller $\eta/s$, which corresponds to a larger partonic scattering cross section, used in the pNJL transport model than that used in the NJL transport model is because partons in the pNJL model are affected by a more attractive scalar potential as shown in Fig.~\ref{fig3}. Although the $\eta/s$ used in the pNJL transport model is smaller than its theoretical lower bound~\cite{Kov05}, it is an effective one since we have neglected its dependence on temperature and chemical potential.

\begin{figure}[t]
    \includegraphics[scale=0.5]{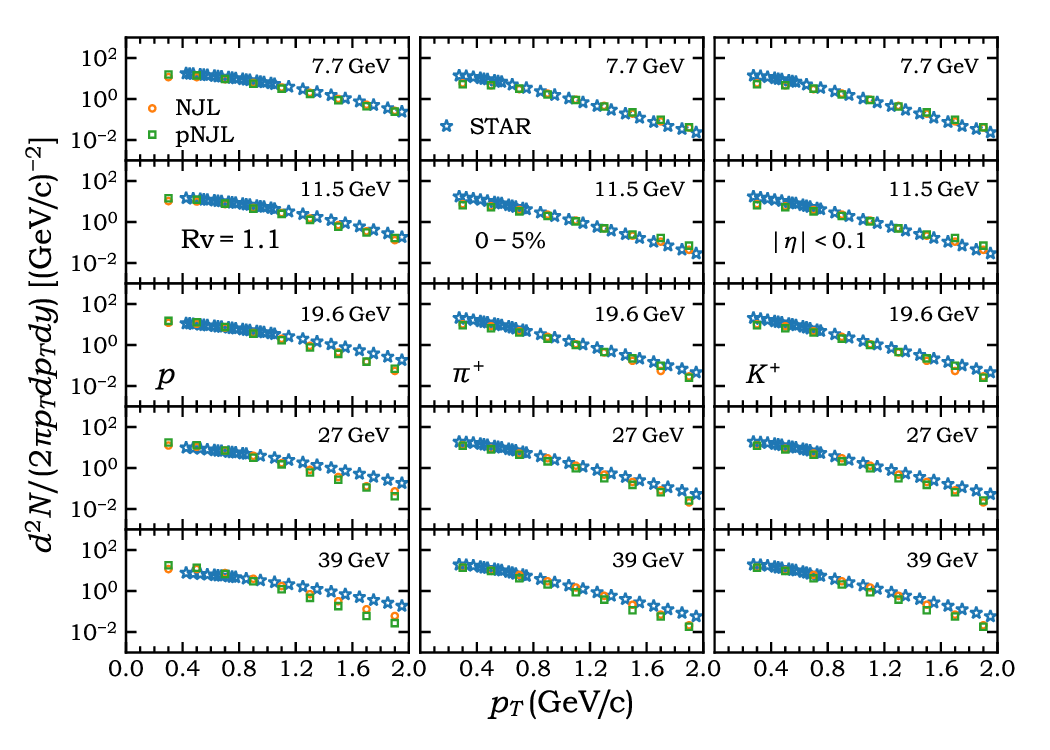}
	\caption{(Color online) Transverse momentum spectra of mid-presudorapidity $\pi^+$, $K^+$, and protons in central ($0-5\%$) Au+Au collisions at different collision energies from the extended AMPT model with the partonic phase described by the NJL or the pNJL model. Experimental data are taken from Ref.~\cite{STAR17}.} \label{fig10}
\end{figure}

In Fig.~\ref{fig10}, we compare the $p_T$ spectra of representative particles, i.e., $\pi^+$, $K^+$, and protons, at different collision energies with corresponding experimental data from the STAR Collaboration~\cite{STAR17}. Since most hadrons are produced in their resonance states right after hadronization, these hadrons experience hadronic elastic and inelastic scatterings and decays during the hadronic evolution. Except for a slightly stiffer $p_T$ spectra for $\pi^+$ at lower collision energies and a softer $p_T$ spectra for protons at higher collision energies, the extended AMPT model describes reasonably well the experimental data from heavy-ion collisions at RHIC-BES energies.

\begin{figure}[h]
	\includegraphics[scale=0.5]{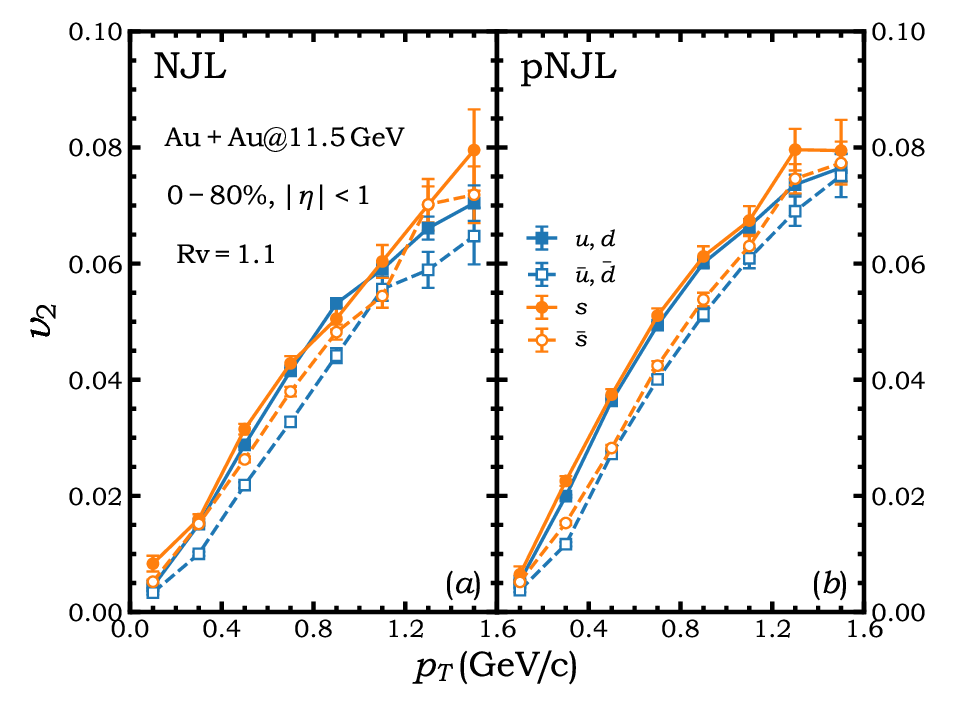}
	\caption{(Color online) Elliptic flows of mid-pseudorapidity quarks and antiquarks as a function of transverse momentum just before hadronization in heavy-ion collisions at $\sqrt{s_{NN}}=11.5$ GeV from both NJL (left) and pNJL (right) transport models.} \label{fig11}
\end{figure}

Before discussing the $v_2$ splitting of final hadrons, we first show in Fig.~\ref{fig11} the $p_T$ dependence of quark and antiquark $v_2$ at the end of the partonic evolution at $\sqrt{s_{NN}}=11.5$ GeV. In the presence of a strong vector interaction with $R_V=1.1$, a larger $v_2$ is observed for quarks than their antiquarks, consistent with the intuitive picture discussed in the beginning of this section. Although the vector potential is slightly stronger in the pNJL transport model than in the NJL model as shown in Fig.~\ref{fig5}, the $v_2$ splittings between light quarks and their antiquarks are similar in the two models, and this is because there are more successful scatterings in the pNJL model than in the NJL model as shown in Fig.~\ref{fig7}. In general, the $v_2$ splitting between high-momentum particles and antiparticles is larger in the pNJL transport model than that in the NJL transport model, since they are not affected by the vector potential due to the momentum cutoff in the latter case. This is especially so for strange quarks with a larger dynamical mass.

\begin{figure}[h]
	\includegraphics[scale=0.6]{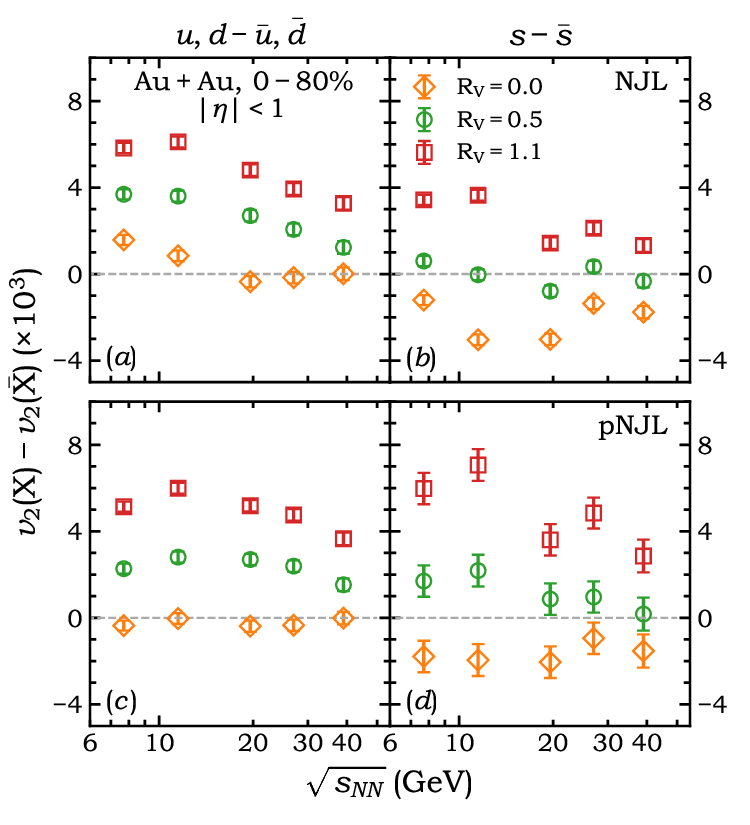}
	\caption{(Color online) Difference between the $p_T-$integrated $v_2$ of quarks and antiquarks just before hadronization for collisions at different energies from both NJL (upper) and pNJL (lower) transport models.} \label{fig12}
\end{figure}

The splitting between $p_T-$integrated $v_2$ of quarks and their antiquarks, which are obtained by their respective average $v_2$ over all $p_T$, at the end of the partonic phase at different collision energies are displayed in Fig.~\ref{fig12}. The $v_2$ splitting is seen to increase with increasing $R_V$. The negative $v_2$ splitting between $s$ and $\bar{s}$ in the case without the vector interaction shown in the right panels of Fig.~\ref{fig12} is due to more successful scatterings for $\bar{s}$ than $s$, as a result of a more diffusive momentum distribution for Pauli blockings for $\bar{s}$. For the case using a larger $R_V$, the $v_2$ splitting shows a non-monotonic dependence on the collision energy, especially for that between light quarks and their antiquarks with a peak $v_2$ splitting at $\sqrt{s_{NN}}=11.5$ GeV. This behavior is similar to the non-monotonic energy dependence of the magnitude of the vector potential shown in Fig.~\ref{fig5}, and this is due to the competition between the collision energy dependence of various factors, e.g., the longitudinal Lorentz contraction, the net quark density, and the expansion speed of the produced matter. This behavior is different from that observed in Ref.~\cite{Xu16}, where the finite thickness of the quark matter in the longitudinal direction was not properly incorporated. Although the $v_2$ splitting between light quarks and their antiquarks is similar in the NJL and pNJL transport models, that between strange quarks and their antiquarks is larger in the pNJL transport model compared with that in the NJL transport model for the reason already given in the discussions on the splitting of differential $v_2$ shown in Fig.~\ref{fig11}.

\begin{figure}[h]
	\includegraphics[scale=0.5]{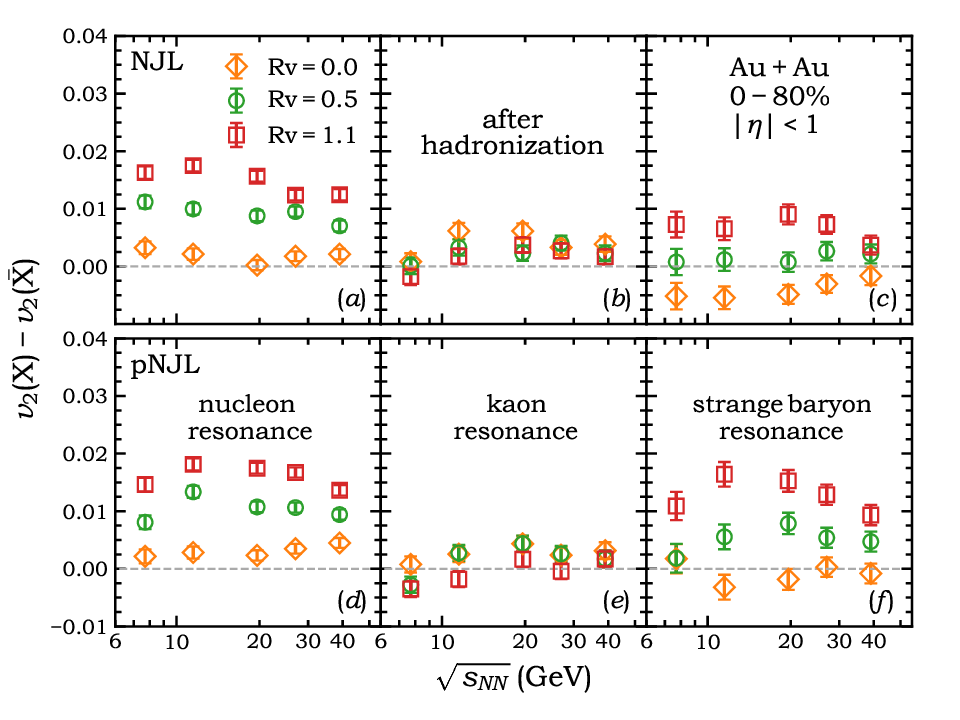}
	\caption{(Color online) Difference between the $p_T-$integrated $v_2$ of nucleon resonances and their antiparticles, positively-charged kaon-like mesons and negatively-charged kaon-like mesons, and strange baryon resonances and their antiparticles, right after hadronization at different collision energies from the extended AMPT model with the partonic phase described by the NJL (upper) or the pNJL (lower) transport model with different values for the reduced vector coupling constant $R_V$.} \label{fig13}
\end{figure}

To see how the $v_2$ splittings between quarks and antiquarks are carried over to those between hadrons and their antiparticles, we display in Fig.~\ref{fig13} the $v_2$ splitting between nucleon resonances and their antiparticles, positively-charged kaon-like mesons ($K^+$ and $K^{*+}$) and negatively-charged kaon-like mesons ($K^-$ and $K^{*-}$), and strange baryon resonances and their antiparticles, right after hadronization when most hadrons are in their resonance states. It is seen that the $v_2$ splitting between baryons and antibaryons first increases and then decreases with the increasing collision energy, consistent with the collision energy dependence of the $v_2$ splitting between quarks and antiquarks shown in Fig.~\ref{fig12}. The situation is more complicated for mesons, which are composed of a quark and an antiquark, due to the opposite effects from the vector potential on quark and antiquark $v_2$. For the larger $R_V=1.1$, although negatively-charged kaon-like mesons have a slightly larger $v_2$ than that of positively-charged ones at low collision energies due to the dominating effect of heavier strange quarks, their $v_2$ splitting is reversed at high collision energies. The $v_2$ splitting between nucleon resonances and their antiparticles as well as that between kaon-like mesons and their antiparticles are seen to be similar in both the NJL and pNJL transport models, while that between strange baryon resonances and their antiparticles is larger in the pNJL transport model than in the NJL transport model, consistent with the larger $v_2(s)-v_2(\bar{s})$ in the pNJL transport model shown in Fig.~\ref{fig12}.

\begin{figure}[h]
	\includegraphics[scale=0.5]{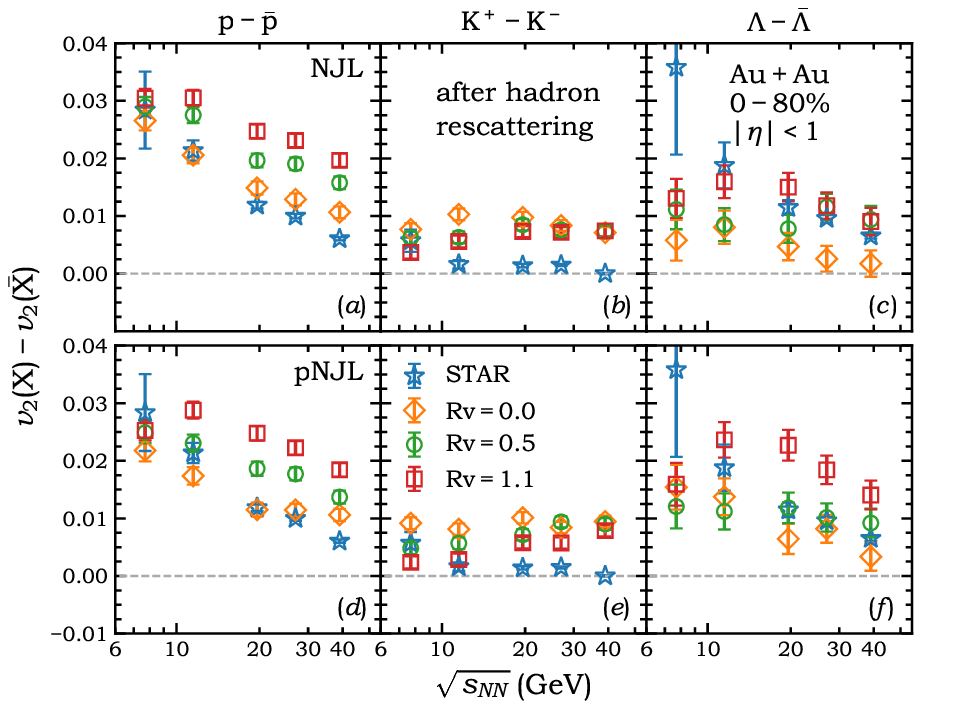}
	\caption{(Color online) Difference between the $p_T-$integrated $v_2$ of freeze-out nucleons and antinucleons, $K^+$ and $K^-$, and $\Lambda$ and $\bar{\Lambda}$ at different collision energies from the extended AMPT model with the partonic phase described by the NJL (upper) or the pNJL (lower) model. Corresponding experimental data are taken from the STAR Collaboration~\cite{STAR14}.} \label{fig14}
\end{figure}

We finally compare in Fig.~\ref{fig14} the $v_2$ splittings between final nucleons and antinucleons~{\footnote{Experimental data are for the $v_2$ splitting between protons and antiprotons, which are compared with that between nucleons and antinucleons from the model calculation for better statistics.}, $K^+$ and $K^-$, and $\Lambda$ and $\bar{\Lambda}$ at different collision energies with  corresponding experimental data from the STAR Collaboration~\cite{STAR14}. It is seen that the $v_2$ splittings between nucleons and antinucleons as well as those between $\Lambda$ and $\bar{\Lambda}$ after hadronic evolution are similar to those of baryons and antibaryons right after hadronization shown in Fig.~\ref{fig13}, and this is because the more attractive potentials for antibaryons than baryons~\cite{Xu12}. This is very different for the $v_2$ splitting between $K^+$ and $K^-$, which has an opposite sign compared to that between positively-charged and negatively-charged kaon resonances shown in the middle panel of Fig.~\ref{fig13}, as a result of the repulsive potential for $K^+$ and the attractive potential for $K^-$~\cite{Xu12}. Even for $R_V=0$, there are $v_2$ splittings between hadrons and their antiparticles from our model calculations, and this is mainly due to their different hadronic potentials. The collision energy dependence of the $v_2$ splittings between hadrons and their antiparticles displayed in Fig.~\ref{fig14} shows a similar non-monotonic behavior as that of the parton $v_2$ splittings, especially for the $v_2$ splitting between $\Lambda$ and $\bar{\Lambda}$. Compared to the experimental data from the STAR Collaboration, the $v_2$ splitting between nucleons and antinucleons can be reproduced reasonably well within $R_V=0 \sim 1.1$ but favors a smaller $R_V$ at higher collision energies. Although our models reproduce the $v_2$ splitting between $K^+$ and $K^-$ at $\sqrt{s_{NN}}=7.7$ GeV, they overestimate their $v_2$ splitting at higher collision energies. Since the partonic and hadronic phase have different effects on the $v_2$ splitting between $K^+$ and $K^-$, a more accurate handling of the lifetimes for the two phases is called for. For the $v_2$ splitting between $\Lambda$ and $\bar{\Lambda}$, our results underestimate the experimental data at $\sqrt{s_{NN}}=7.7$ GeV, while the data at higher collision energies can be reproduced by the extended AMPT model with $R_V=1.1$ in the NJL transport model and $R_V=0.5$ in the pNJL transport model. A detailed fit of the value of $R_V$ by taking into account its collision energy dependence via the incorporation of a baryon chemical potential dependent $R_V$ needs to be investigated in future studies.

\section{Conclusion}
\label{summary}

With the partonic phase of heavy-ion collisions described by a partonic transport model that is based on the 3-flavor Nambu-Jona-Lasinio (NJL) model or its extension with the inclusion of the Polyakov loops, we have revisited the elliptic flow ($v_2$) splittings between particles and their antiparticles in relativistic heavy-ion collisions at RHIC-BES energies. We have checked that for a partonic matter in a box with periodic boundary conditions, the NJL and pNJL transport models lead to their respective thermal momentum distributions if the Pauli-blocking factors are included in the collision integral of the NJL model and a properly modified treatment of parton scatterings is introduced in the pNJL transport model. In a partonic matter of a given temperature and baryon chemical potential, it is found that the pNJL transport model gives a higher net quark density and a stronger vector potential than those from the NJL transport model.  As a result, a larger $v_2$ splitting is found between $s$ and $\bar{s}$ quarks in heavy-ion collisions, and this affects the $v_2$ splitting between final $\Lambda$ and $\bar{\Lambda}$. On the other hand, due to the competition between effects from the longitudinal Lorentz contraction, the baryon chemical potential, and the expansion time at different collision energies, the magnitude of the vector potential and the $v_2$ splitting between quarks and antiquark changes non-monotonically with the increasing collision energy for a given vector coupling constant in both the NJL and the pNJL model. Consequently, the $v_2$ splitting between final hadrons and their antiparticles also shows a non-monotonic dependence on the collision energy. The resulting $v_2$ splitting between $K^+$ and $K^-$ reproduces the STAR data at $\sqrt{s_{NN}}=7.7$ GeV but overestimates it at higher collision energies. For the $v_2$ splitting between protons and antiprotons, the STAR data at $\sqrt{s_{NN}}=7.7$ GeV can be reproduced with the range of vector coupling in both NJL and pNJL transport models considered in the present study, while the STAR data at higher collision energies favor a weak vector interaction in these transport models. Although our calculations underestimate the STAR data on the $v_2$ splitting between $\Lambda$ and $\bar\Lambda$ at $\sqrt{s_{NN}}=7.7$ GeV, a strong (weak) vector interaction is favored in the NJL (pNJL) transport model for collisions at higher energies. Our study has therefore provided useful insight on the determination of the vector interaction in the partonic matter produced in heavy-ion collisions, which will help understand the equation of state of quark matter at large baryon chemical potentials and thus the QCD phase diagram.

\begin{acknowledgments}

JX was supported by the National Natural Science Foundation of China under Grant No. 11922514. CMK was supported by
the US Department of Energy under Award No. DESC0015266 and the Welch Foundation under Grant No. A-1358.

\end{acknowledgments}

\begin{appendix}
	
\section{Relation between the parton scattering cross section and the specific shear viscosity}
\label{xsection}

The specific shear viscosity in the NJL model has been studied in Ref.~\cite{Gho16} from the quark-meson interactions, and that in the pNJL model has been studied in Ref.~\cite{Sol21} based on the relaxation time approximation. In the present study, we determine the isotropic parton scattering cross section $\sigma$ from the shear viscosity $\eta$ through the empirical relation
\begin{equation}
\eta = \frac{4\left\langle p\right\rangle}{15\sigma_{tr}}, \label{eta}\\
\end{equation}
where
\begin{equation}
\left\langle p\right\rangle = \frac{2N_c\sum_{q} \int \frac{\mathrm{d}^3p}{\left(2\pi\right)^3} p (n_q+\bar{n}_q)}{2N_c\sum_{q} \int \frac{\mathrm{d}^3p}{\left(2\pi\right)^3} (n_q+\bar{n}_q)}
\end{equation}
is the average parton momentum with $n_q$ and $\bar{n}_q$ being the occupation probability for parton species $q=u$, $d$, $s$ quarks and their antiquarks, i.e., $n_q=f_q$ and $\bar{n}_q=\bar{f_q}$ for the NJL model and $n_q=F_q$ and $\bar{n}_q=\bar{F_q}$ for the pNJL model. The $\sigma_{tr}$ in Eq.~(\ref{eta}) is the transport cross section defined as
\begin{equation}
\sigma_{tr} = \int\mathrm{d}\Omega\frac{\mathrm{d}\sigma}{\mathrm{d}\Omega}\left(1-\cos^2\theta\right). \label{sigma_tr}
\end{equation}
For an isotropic cross section $\sigma$, one has $\sigma_{tr}=\frac{2}{3}\sigma$. The entropy density $s$ of a quark matter is given by
\begin{eqnarray}
s &=& -2N_c \sum_{q} \int\frac{\mathrm{d}^3p}{\left(2\pi\right)^3}\left[n_q \ln  n_q+\left(1-n_q\right)\ln\left(1-n_q\right)\right] \notag\\
 &-& 2N_c \sum_{q} \int\frac{\mathrm{d}^3p}{\left(2\pi\right)^3}\left[\bar{n}_q \ln  \bar{n}_q +\left(1-\bar{n}_q \right)\ln\left(1-\bar{n}_q \right)\right]. \notag\\\label{sden}
\end{eqnarray}

\begin{figure}[ht]
	\includegraphics[scale=0.5]{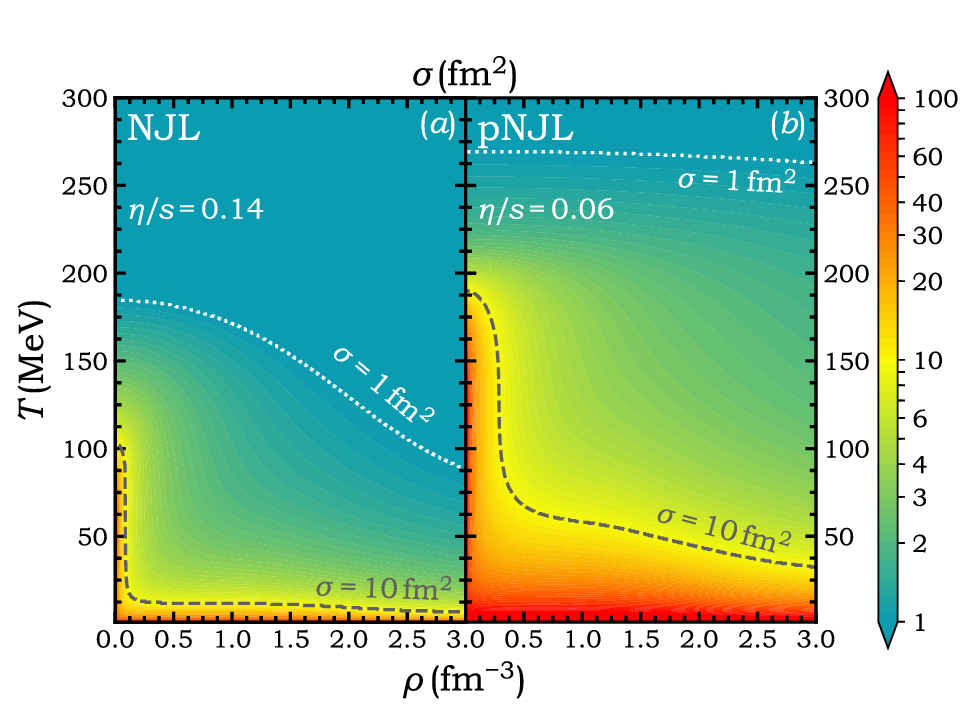}
	\caption{(Color online) The $\rho$ and $T$ dependence of parton scattering cross section determined from a constant specific shear viscosity $\eta/s=0.14$ for the NJL (left) and 0.06 for the pNJL (right) model. } \label{fig15}
\end{figure}

By using a constant specific shear viscosity $\eta/s=0.14$ for the NJL model and 0.06 for the pNJL model, the cross section $\sigma$ at different net quark number densities $\rho$ and temperatures $T$ are displayed in Fig.~\ref{fig15}. It is seen that the cross section is around 1 fm$^2$ in the high-temperature region, while it can be as large as 10 fm$^2$ at low temperatures and/or densities.

\section{Stochastic treatment of parton scattering with Pauli blockings}
\label{stochastic}

We employ the stochastic method~\cite{Xu05} to treat elastic scatterings between partons. The collision probability for a pair of partons in a volume $\Delta V=\Delta x \Delta y \Delta z$ and a time interval $\Delta t$ is
\begin{equation}\label{sto}
P_c = v_{12}\sigma\frac{\Delta t}{\Delta V},
\end{equation}
with
\begin{equation}
v_{12}=\frac{\sqrt{[s-(M_1+M_2)^2][s-(M_1-M_2)^2]}}{2E_1E_2}
\end{equation}
being the M$\phi$ller velocity. In the above, $s$ is the square of the invariant mass of the parton pair, and $E_{1(2)}=\sqrt{M_{1(2)}^2+p_{1(2)}^2}$ is the energy of the parton, with $M_{1(2)}$ being the in-medium parton mass from the NJL or the pNJL model, and $p_{1(2)}$ being the momentum of the parton in the rest frame of the cell. The $\sigma$ in Eq.~(\ref{sto}) is an isotropic cross section, determined by the specific shear viscosity as detailed in Appendix~\ref{xsection}. The volume of each cell is set as $\Delta x = \Delta y = 0.5$ fm and $\Delta z = 0.25$ fm, and only parton pairs in the same cell can collide with each other. The time step $\Delta t=0.02\,\mathrm{fm/c}$ is used in the calculation. The case $P_c>1$ is equal to the case of $P_c=1$, and in this way the collision rate from a too large cross section is limited by the values of $\Delta V$ and $\Delta t$.

After an attempted collision, the momentum of each parton in the center-of-mass frame of the collision pair is sampled isotropically. After the momenta of the two partons are boosted to the rest frame of the local cell, they change from $p_{1(2)}$ to $p_{1(2)}^\prime$. For a given flavor, the occupation probability in the rest frame of the local cell after the attempted collision is calculated from the test-particle method,
\begin{equation}\label{TP}
n_q(\vec{r},\vec{p}) \sim \frac{1}{N_{TP}}\sum_{i \in q} \delta(\vec{r}-\vec{r}_i)\delta(\vec{p}-\vec{p}_i^\prime),
\end{equation}
where $N_{TP}$ is the test particle number. The occupation probability is calculated separately for partons with different flavors and also separately for quarks and antiquarks. The Pauli blocking probability is then $1-[1-n_q(\vec{r},\vec{p}_1^\prime)][1-n_q(\vec{r},\vec{p}_2^\prime)]$. The collision is successful if a random number within $[0,1]$ is larger than the Pauli blocking probability, otherwise the momenta of the two partons $p_{1(2)}$ are retained.

\section{Modified Pauli blocking in the pNJL transport model}
\label{collisions}

The collision integral in the Boltzmann equation with quantum correction generally contains the factor ($f_3 f_4 g_1 g_2 - f_1 f_2 g_3 g_4 $) for the $1+2 \rightarrow 3+4$ process, where $g_{1,2,3,4}$ is equal to $1-f_{1,2,3,4}$ if the Fermi-Dirac statistics is adopted. For the NJL transport model, as mentioned in Appendix B, the successful collision probability is $(1-f_3)(1-f_4)$ when the Pauli blocking is applied. For the Fermi-Dirac distribution given in Eq.~(\ref{fqnjl}), $g=\xi f$ with $\xi = e^{(E-\tilde{\mu})/T}$ (neglecting the subscript flavor index $q$) is satisfied. To obtain the thermal distribution $F$ [Eq.~(\ref{Fq})] in the pNJL transport model, a similar consideration can be applied by taking the successful collision probability as $g_3 g_4$ with $g = \xi F$. Although $F_{3,4}$ can be calculated using the test-particle method through Eq.~(\ref{phase-space}), $\xi$ is a thermodynamic quantity, which needs to be calculated by assuming local thermal equilibrium as described below.

Taking the expression of $F=F_q$ as an example, Eq.~(\ref{Fq}) can be rewritten as
\begin{equation}
\xi^3 + A \xi^2 + B \xi + C = 0, \label{Feq}
\end{equation}
with
\begin{equation}
A=\frac{\Phi (3F-1)}{F},~B=\frac{\bar{\Phi}(3F-2)}{F},~C=\frac{F-1}{F}.
\end{equation}
For $0<F<1$, the only real solution to Eq.~(\ref{Feq}) is
\begin{eqnarray}
\xi &=& \frac{1}{3} \left[-A+(-D+\sqrt{D^2+H^3})^{1/3}\right. \notag\\
&+&\left.(-D-\sqrt{D^2+H^3})^{1/3}\right],
\end{eqnarray}
with
\begin{equation}
D = \frac{27C-9AB+2A^3}{2},~H=3B-A^2.
\end{equation}
In this way, $\xi$ can be expressed as a function of $F$, $\Phi$, and $\bar{\Phi}$. Similarly, the $\xi^\prime$ in ${\bar F}_q$ can be expressed as a function of $\bar{F}$, $\Phi$, and $\bar{\Phi}$.
\end{appendix}

\end{document}